\documentclass[aps,pra,twocolumn,showpacs]{revtex4}

\usepackage{amssymb}
\usepackage{amsmath}
\usepackage{graphicx}

\begin{document}

\title{Long-range entanglement generation via frequent measurements}
\author{L.-A. Wu, D.A. Lidar, S. Schneider}
\affiliation{Chemical Physics Theory Group, Chemistry Department, University of Toronto,
80 St. George Street, Toronto, Ontario M5S 3H6, Canada}

\begin{abstract}
A method is introduced whereby two non-interacting quantum subsystems, that
each interact with a third subsystem, are entangled via repeated projective
measurements of the state of the third subsystem. A variety of physical
examples are presented. The method can be used to establish long range
entanglement between distant parties in one parallel measurement step, thus
obviating the need for entanglement swapping.
\end{abstract}

\maketitle

\section{Introduction}

Entanglement is at the heart of quantum information processing \cite{Gruska:book}, and is a resource that allows quantum processes to outperform
their classical counterparts for tasks such as computation \cite{Ekert:98},
communication \cite{Raz:99}, and cryptography \cite{Ekert:91}. The standard
way to generate entanglement between distinguishable and initially separable
particles is to let them interact directly for a certain amount of time:
any non-trivial two-body Hamiltonian is capable of generating entanglement
in this manner \cite{Zhang:03}. However, the direct interaction method poses
limitations in the context of the generation of \emph{long-range}
entanglement, since in many systems the interaction strength typically
decreases at least as fast as some power of the distance between particles.
Here we introduce a different paradigm for entanglement generation: we show
that it is possible \emph{to entangle two particles that never interact
directly by means of repeated measurements of a third subsystem that interacts with both}. In addition to its
conceptual interest, we show that this scheme offers practical advantages for long-range
entanglement generation.

We remark that other alternatives to the
direct interaction method for entanglement generation are well known. E.g.,
in \cite{Knill:00} linear optics is augmented by \emph{measurements} to
generate entanglement and perform quantum computation. Measurements of the
phase of light transmitted through a cavity have also been proposed in \cite{Sorensen:03} as a method to prepare entangled states and implement quantum
computation in the case of atoms in optical cavities. \emph{Decoherence} can
also be used to create entanglement: e.g., atoms in a leaky cavity can
become entangled conditioned on null detection at a photo detector placed
outside the cavity \cite{Plenio:99,Beige:00a}. In Ref.~\cite{Cabrillo:99}
entanglement is generated in a similar manner conditioned instead upon
detection.

Here we present a rather general theoretical framework for
measurement-generated entanglement. Our scheme is inspired by the recent
work by Nakazato, Takazawa, and Yuasa (NTY) \cite{Nakazato:03}, who
investigated the effects of repeated rapid projective measurements on one
subsystem of a bipartite system. Provided that the resulting operator on the
unobserved subsystem has a non-degenerate largest eigenvalue, NTY showed that \emph{the unobserved subsystem is gradually
projected into a pure state, independent of the initial state}. This result
leads to the question of whether a similar measurement scheme is capable of
generating entanglement when the unobserved subsystem is itself
multi-partite, non-interacting, and in an arbitrary initial state. We answer
this question in the affirmative for a wide range of model systems, and
establish general conditions for validity of the method. We note that an
additional advantage of this measurement-based method for generating
entanglement over the direct-interaction based method is that does not
depend on sensitive timing of interactions: instead, entanglement is
gradually purified as the number of measurements increases, and (under
appropriate conditions) can be made arbitrarily high. An important
conclusion from our study is that the method can be used to establish long
range entanglement between distant particles, by measuring along a chain of
intermediate particles. This provides an alternative to entanglement
swapping that does not scale with the chain length. A similar result
-- long range entanglement from local (but non-repeated) measurements -- was obtained in
Ref.~\cite{Verstraete:04} for the ground state of an antiferromagnetic spin chain. We now turn a detailed
description of our method.

\section{Preliminaries}

NTY considered the following scenario: Consider a bi-partite system composed
of subsystems $A,B$, initially in the separable state $\rho _{AB}(0)=|\phi
\rangle \langle \phi |\otimes \rho _{B}$, where $A$ is in a pure state $%
|\phi \rangle $ and $\rho _{B}$ is arbitrary. Subsystem $A$ is subject to
projective measurements $P_{A}=|\phi \rangle \langle \phi |$ applied with
period $\tau $. In between measurements the system evolves under the
Hamiltonian 
\begin{equation}
H=H_{A}+H_{B}+H_{AB}
\end{equation}
(resp., the sum of free Hamiltonians and an interaction). It can be shown 
\cite{Nakazato:03} that after $M$ such measurements the state of subsystem $B
$, \emph{given} that all outcomes were $|\phi \rangle $, is 
\begin{equation}
\rho _{B}(M\tau )=\frac{V_{B}(\tau )^{M}\rho _{B}V_{B}^{\dagger }(\tau )^{M}%
}{P_{M}},  \label{eq:rhoB}
\end{equation}%
where $P_{M}$ is the survival probability, i.e., the probability of finding
subsystem $A$ in its initial state,%
\begin{equation}
P_{M}=\mathrm{Tr}_{B}[V_{B}(\tau )^{M}\rho _{B}V_{B}^{\dagger }(\tau )^{M}]
\label{successprob}
\end{equation}%
where $U=\exp (-i\tau H)$ (we use units where $\hbar =1$), and where 
\begin{equation}
V_{B}(\tau )\equiv \langle \phi |U|\phi \rangle   \label{eq:VB}
\end{equation}%
is an operator on subsystem $B$ that is in general \emph{not} Hermitian.
However, one may still find left- and right-eigenvectors with complex
eigenvalues, whose modulus can be shown to be bounded between $0$ and $1$.
The central result of NTY is that in the limit of large $M$ and small but
finite $\tau $, $\rho _{B}(M\tau )$ \emph{tends to a pure state independent
of }$B$\emph{'s initial state}. This result assumes that the largest
eigenvalue $\lambda _{\mathrm{\max }}$ of $V_{B}(\tau )$ is non-degenerate.
The final pure state $\rho _{B}(M\tau )$ is then the corresponding right
eigenvector $|u_{\max }\rangle $ and this outcome is found with probability 
\begin{equation}
P_{M}\rightarrow |\lambda _{\mathrm{\max }}|^{2}\langle u_{\max }|u_{\max
}\rangle \langle v_{\max }|\rho _{B}|v_{\max }\rangle ,
\end{equation}
where $|v_{\max }\rangle $ is the corresponding left eigenvector. Note that
because of $\tau $'s finiteness the dynamics here is distinct from the Zeno
effect and \textquotedblleft quantum Zeno dynamics\textquotedblright\ \cite%
{Facchi:PRL02}.

\begin{figure}[tbp]
\includegraphics[width=7cm]{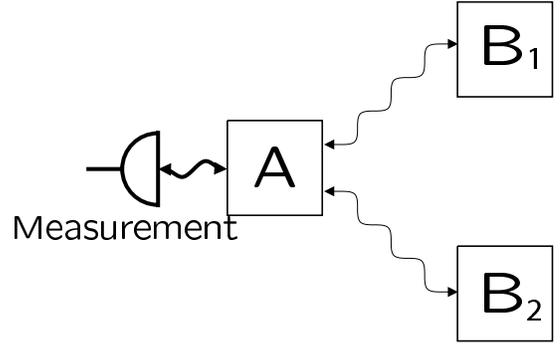}
\caption{Schematic illustration of our model: Systems $B_1$ and $B_2$ are
coupled to system $A$; system $A$ can be measured with a projective
measurement. Note that no direct coupling exists between systems $B_1$ and $B_2$.}
\label{figone}
\end{figure}

We now extend NTY's model by allowing $B$ itself to be composed of multiple 
\emph{non-interacting} subsystems, $B=\{B_{1},B_{2},...,B_{N}\}$, and pose
the question: is the NTY measurement procedure capable of generating
entanglement amongst $B$'s subsystems, assuming that they all interact with
the \textquotedblleft station\textquotedblright\ $A$? The situation is
illustrated in Fig.~\ref{figone}. To answer this questions we consider the
following model:\ We set all internal Hamiltonians $H_{A}=H_{B_{i}}=0$, so
that $H=H_{AB}$ (equivalently, we can always transform to an interaction
picture rotating with the internal Hamiltonians; this will make $H_{AB}$
time dependent, but it will not introduce couplings between subsystems $%
B_{i} $). We assume that the $B$ subsystems are all qubits and the
interaction between $A$ and all the $B_{i}$ is identical. We can then define
total quasi-spin operators $\sigma _{\beta }\equiv \sum_{i=1}^{N}\sigma
_{\beta }^{i}$ and write the interaction as $H_{AB}=\sum_{\alpha \in
\{x,y,z\}}h_{\alpha }A_{\alpha }\otimes \sigma _{\alpha }$ ($h_{\alpha }$
real) or, more simply, 
\begin{equation}
H_{AB}=\overrightarrow{A}\cdot \overrightarrow{\sigma }  \label{eq3-1}
\end{equation}%
where the parameters $h_{\alpha }$ are included in the vector $%
\overrightarrow{A}$. We do not restrict subsystem $A$.

\section{General theory}

\subsection{A good basis}

We construct a basis that block-diagonalizes the effective post-measurement
evolution operator $V_{B}(\tau )$. Let a basis for the $A$ subsystem be the
measurement state $\left\vert \phi \right\rangle $ and any set of states
orthonormal to it, denoted $\left\vert \phi _{l}^{\bot }\right\rangle $, $%
l=1,...,d-1$. Let $N$ be even and let a basis for the $B$ subsystem be
constructed out of the usual spin basis $|S,M_{S}\rangle $, where $S$ is the
total quasi-spin of the $N$ particles and $M_{S}$ its projection along the $%
z $-axis. We denote the (orthonormal) singlet states by $|s_{j}\rangle $, $%
j=1,...,D_{N}=N!/[(N/2+1)!(N/2)!]$, and the remaining (orthonormal) states
with $S>0$ by $|t_{k}\rangle $, $k=1,...,K_{N}$, $K_{N}=2^{N}-D_{N}$.

\textit{Proposition 1}:\ Consider the ordered basis $\{\left\vert \phi
\right\rangle ,\left\vert \phi _{1}^{\bot }\right\rangle ,...,\left\vert
\phi _{d-1}^{\bot }\right\rangle \}\otimes \{|s_{1}\rangle
,...,|s_{D_{N}}\rangle ,|t_{1}\rangle ,...,|t_{K_{N}}\rangle \}$. In this
basis we have the block-diagonal representation 
\begin{equation}
V_{B}(\tau )=\left( 
\begin{tabular}{l|l}
$I_{D_{N}}$ & $\mathbf{0}$ \\ \hline
$\mathbf{0}$ & $V_{B}^{s}$%
\end{tabular}%
\right) ,  \label{eq:VB-blockrep}
\end{equation}%
where $I_{D_{N}}$ is a $D_{N}\times D_{N}$-dimensional identity matrix. The
maximal eigenvalue of $V_{B}(\tau )$ is $1$, and is at least $D_{N}$-fold
degenerate.

\textit{Proof}: The singlets states, $|S,M_{S}\rangle =|0,0\rangle $, are
annihilated by the $\sigma _{\beta }$ when $N$ is even (since they are
states with zero total quasi-spin). Therefore $H_{AB}|s_{j}\rangle =0$,
independent of the state of subsystem $A$. In the ordered basis $H_{AB}$ is
thus represented as%
\begin{equation}
H_{AB}=\left( 
\begin{tabular}{l|l}
$\mathbf{0}$ & $\mathbf{0}$ \\ \hline
$\mathbf{0}$ & $H_{AB}^{^{\prime }}$%
\end{tabular}%
\right)
\end{equation}%
where the dimension of the upper-left block is $D_{N}\times D_{N}$ and that
of the lower-right block is $K_{N}\times K_{N}$. Then%
\begin{equation}
U_{AB}=e^{-i\tau H_{AB}}=\left( 
\begin{tabular}{l|l}
$I_{D_{N}}$ & $\mathbf{0}$ \\ \hline
$\mathbf{0}$ & $U_{AB}^{\prime }$%
\end{tabular}%
\right) ,  \label{eq5}
\end{equation}%
where $U_{AB}^{\prime }=W^{\prime }e^{-i\tau \Lambda }W^{\prime \dag }$ is
unitary and $W^{\prime }$ is the unitary matrix that diagonalizes $%
H_{AB}^{\prime }$: $W^{\prime \dag }H_{AB}^{\prime }W^{\prime }=\Lambda =%
\mathrm{diag}(\lambda _{1},...,\lambda _{K_{N}})$. Taking the expectation
value wrt $|\phi \rangle $ we then find:%
\begin{equation}
V_{B}(\tau )=\sum_{j=1}^{D_{N}}|s_{j}\rangle \langle s_{j}|+V_{B}^{s},
\label{eq7}
\end{equation}%
which is the claimed result, with $V_{B}^{s}\equiv \sum_{k,k^{\prime
}=1}^{K_{N}}|t_{k}\rangle \left( U_{AB}^{\prime }\right) _{\phi k,\phi
k^{\prime }}\langle t_{k^{\prime }}|$ a $K_{N}\times K_{N}$-dimensional
matrix. That the maximal eigenvalue is $1$ follows from unitarity of $%
U_{AB}^{\prime }$ \cite{Nakazato:03}, and that it is at least $D_{N}$-fold
degenerate is immediate from the representation (\ref{eq:VB-blockrep}). QED

\textit{Proposition 2 }\cite{Nakazato:03}: Non-degeneracy of $V_{B}(\tau )$
is a necessary condition for obtaining a pure state in the limit of large $M$%
.

\textit{Proof}: In the degenerate case it follows from Eq.~(\ref{eq:rhoB})\
that the method yields an equally weighted sum of degenerate pure states
corresponding to the maximum eigenvalue of $V_{B}(\tau )$. QED

Note that $D_{2}=1$. Hence in the totally symmetric case we have been
considering so far, for $N=2$ \emph{the method will generate a pure
(maximally entangled) singlet state}, provided $V_{B}^{s}$ has maximal
eigenvalue with modulus smaller than $1$. We give examples of corresponding
Hamiltonians below. On the other hand, $D_{4}=2$ and hence due to degeneracy
the method will not produce a pure entangled state for $N\geq 4$. However,
we can still generate pure-state entanglement in the $N=4$ case by breaking
the total permutation symmetry and only preserving the symmetry between $1,2$
and $3,4$ [e.g., having a coupling to subsystem $A$ such that $\sigma
_{\beta }=a_{1}(\sigma _{\beta }^{1}+\sigma _{\beta }^{2})+a_{2}(\sigma
_{\beta }^{3}+\sigma _{\beta }^{4})$, $a_{1}\neq a_{2}\neq 0$]. We can then
project out a \emph{one}-dimensional subspace $|s\rangle _{12}|s\rangle
_{34} $, $|s\rangle _{ij}=(|0_{i}1_{j}\rangle -|1_{i}0_{j}\rangle )/\sqrt{2}$
is the singlet state, which is entangled for $1,2$ and $3,4$ but separable
across the $12:34$ partition. Similarly, for $N=2n\geq 6$ and $\sigma
_{\beta }=\sum_{i=1}^{n}a_{i}(\sigma _{\beta }^{2i-1}+\sigma _{\beta }^{2i})$%
, $a_{i}\neq a_{j}\neq 0$ the projected state is $|s\rangle _{12}|s\rangle
_{34}......|s\rangle _{n-3,n-2}|s\rangle _{n-1,n}$.

\subsection{Invariance}

The results of the above discussion are invariant as long as the
Hamiltonians belong to the family $\{U_{B}H_{AB}U_{B}^{\dagger }\}$, where $%
U_{B}$ is an arbitrary unitary transformation of the Hamiltonian of
subsystem $B$. This enables the method to generate entangled states
equivalent under local transformations. The invariance of $%
H_{AB}|s_{j}\rangle =0$ is $H_{AB}^{\prime }|s_{j}^{\prime }\rangle =0$,
where $H_{AB}^{\prime }=U_{B}H_{AB}U_{B}^{\dagger }$ and $|s_{j}^{\prime
}\rangle =U_{B}|s_{j}\rangle $. Thus, e.g., in the case of $N=2$, we can
generate the other three Bell states by applying $U_{B}=X_{2},Y_{2},Z_{2}$
(where $X_{2}$ is the Pauli matrix $\sigma _{x}$ acting on subsystem $B_{2}$%
, etc.), which results, respectively, in $|s^{\prime }\rangle =\frac{1}{%
\sqrt{2}}(|11\rangle +|00\rangle ),\frac{1}{\sqrt{2}}(|00\rangle -|11\rangle
),\frac{1}{\sqrt{2}}(|01\rangle +|10\rangle )$ as the outcome of the method.

\subsection{Degenerate maximal eigenvalue}

What happens when there are other states, either singlets or coming from $%
V_{B}^{s}$, that also have an eigenvalue with modulus $1$? In this case
NTY's result does not apply and entanglement may or may not be spoiled
(though it is always reduced from maximally entangled). Two examples will
illustrate: (i) $N=2$:\ Suppose there is a triplet state $|00\rangle $ that
also has $|\lambda _{\mathrm{\max }}|=1$. Then the resulting state is the
mixture $|s\rangle \langle s|+|00\rangle \langle 00|$. This state is
entangled as its partial transpose \cite{Peres:96} has a negative eigenvalue
of $-0.207$. We will encounter this case in the Heisenberg model below. (ii) 
$N=4$:\ Suppose the two singlet states $|s_{1}\rangle =|s\rangle
_{12}|s\rangle _{34}$ and $|s_{2}\rangle =\frac{1}{\sqrt{3}}[|t_{+}\rangle
_{12}|t_{-}\rangle _{34}+|t_{-}\rangle _{12}|t_{+}\rangle
_{34}-2|t_{0}\rangle _{12}|t_{0}\rangle _{34}]$ (where $|t_{\alpha }\rangle $
are triplets with projection quantum number $\alpha $) appear with
eigenvalue one, but no other states do. With respect to the $12:34$ cut $%
|s_{1}\rangle $ is a product state but $|s_{2}\rangle $ is clearly
entangled. This state, just like in the previous example, has negative
partial transpose, and we have entanglement \emph{across the }$12:34$\emph{\
cut}. These examples illustrate that degeneracy still allows for \emph{mixed
state entanglement} to be generated by our method. This is useful
entanglement in the sense that it can be used for teleportation and all
other QIP primitives \cite{Horodecki:97}.

\section{Examples}

We now discuss examples, first limiting ourselves to $A$ being a qubit and $%
N=2$. Our task then reduces to calculating the eigenvalues of the $3\times 3$
matrix $V_{B}^{s}=\langle \phi |U_{AB}^{\prime }|\phi \rangle $. However,
this requires diagonalization of $U_{AB}^{\prime }$, an $6\times 6$ matrix,
so cannot be done analytically in complete generality. Our basis for $A$ is $%
\{|\phi \rangle ,|\phi ^{\bot }\rangle \}$ and for $B$ is
\begin{eqnarray}
\vec{\beta}\equiv
\{|s\rangle &=&\frac{1}{\sqrt{2}}(|01\rangle -|10\rangle ),|t_{-}\rangle
=|00\rangle ,\notag \\
|t_{0}\rangle &=&\frac{1}{\sqrt{2}}(|01\rangle +|10\rangle
),|t_{+}\rangle =|11\rangle \}\},
\label{eq:beta}
\end{eqnarray}
without loss of generality (recall the
invariance discussion above).

\subsection{Axial symmetry model}

Suppose $A_{z}=0$ and $A_{x},A_{y}$ satisfy $%
[A_{x},A_{y}^{2}]=[A_{y},A_{x}^{2}]=0$ (or $A_{x}$ or $A_{y}=0),$ whence the
Hamiltonian reads $H=[A_{x}(X_{1}+X_{2})+A_{y}(Y_{1}+Y_{2})]/2$. We first
consider as a special case the $XY$ model: $A_{x}=JX$ and $A_{y}=JY$,
relevant for quantum information processing in solid state
\cite{Mozyrsky:01,Imamoglu:99,Quiroga:99} and atomic \cite{Zheng:00}
  systems. It follows after some algebra
that $H^{3}=\left\vert A\right\vert ^{2}H$, where $\left\vert A\right\vert
^{2}=A_{x}^{2}+A_{y}^{2}$. Therefore, the evolution is 
\begin{equation}
U_{AB}=e^{-i\tau H}=I-2\left( \frac{H}{|A|}\right) ^{2}\sin ^{2}\frac{\tau
\left\vert A\right\vert }{2}-i\frac{H}{|A|}\sin \tau \left\vert A\right\vert
,  \label{eq8}
\end{equation}%
which means that $V_{B}^{s}$ is a function only of $\langle \phi |H|\phi
\rangle $ and $\langle \phi |H^{2}|\phi \rangle $. In particular, for the $%
XY $ model we find, using $\left\vert A\right\vert =\sqrt{2}J$,

\begin{widetext}
\begin{eqnarray*}
V_{B}(\tau ) &=&\cos ^{2}\frac{\tau J}{\sqrt{2}}-\frac{%
X_{1}X_{2}+Y_{1}Y_{2}-\langle Z\rangle (Z_{1}+Z_{2})}{2}
\sin ^{2}\frac{\tau J}{\sqrt{2}}-i\frac{\langle X\rangle
(X_{1}+X_{2})+\langle Y\rangle (Y_{1}+Y_{2})}{2\sqrt{2}}\sin \sqrt{2}\tau J
\end{eqnarray*}%
\end{widetext}
where $\langle X\rangle \equiv \langle \phi |X|\phi \rangle $, etc. It is
simple to check that, as required from our general result, $V_{B}|s\rangle
=|s\rangle $, i.e., the singlet state has eigenvalue $1$. Whether this
eigenvalue is degenerate is now seen to depend on the measurement of the $A$
subsystem: If we choose $|\phi \rangle $ to be a $\sigma _{z}$-eigenstate
then the additional eigenvalues of $V_{B}$ are found to be $\{1,\cos \sqrt{2}%
\tau J,\cos \sqrt{2}\tau J\}$, while if we choose $|\phi \rangle $ to be a $%
\sigma _{x}$ or $\sigma _{y}$-eigenstate then we find that the additional
eigenvalues are $\{\cos ^{2}q,1+3\cos 2q\pm \sqrt{\cos 4q+\sin ^{4}q-1}\}$,
where $q=\frac{\tau J}{\sqrt{2}}$. Thus, if we measure along $z$ then there
is never pure state entanglement, but at times $\tau $ other than $n\pi /%
\sqrt{2}J$ we have a mixed entangled state, since the eigenstate
corresponding to the additional eigenvalue $1$ is found to be $|11\rangle $
(and recall the discussion above). In the case of measurement along $x$ or $%
y $ the eigenvalues periodically have modulus $1$, and with the exception of
those times degenerary is avoided and we do obtain a pure entangled state
(in particular, it is simple to show that this is true for all $\tau <\sqrt{2%
}\arccos (1/3)/J$).

\begin{figure}[tbp]
\includegraphics[width=7cm]{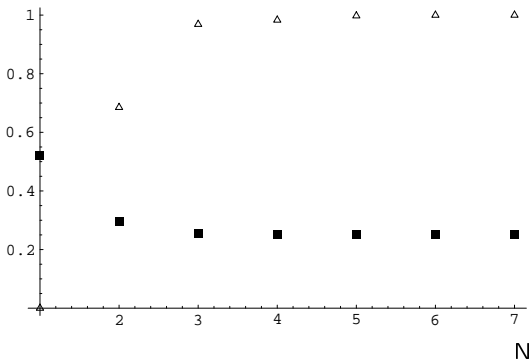}
\caption{Concurrence ($\triangle $) and success probability ($\blacksquare $%
) $P_{M}$ for $|\protect\phi \rangle =\frac{1}{\protect\sqrt{2}}(|0\rangle
+|1\rangle )$ and $\protect\tau J=\protect\pi /2$.}
\label{fig2}
\end{figure}

Next we consider the performance of the scheme after a \emph{finite} number
of steps (Fig.~\ref{fig2}). To do so we calculate the concurrence \cite%
{Wootters:98} after $M$ steps of evolution/measurement for $J\tau =\pi /2<%
\sqrt{2}\arccos (1/3)$, in the case of $|\phi \rangle $ a $\sigma _{x}$
eigenstate. In this case, as shown above, the singlet state is generated by
the scheme. We also plot the survival probability $P_{M}$ as given by Eq.~(%
\ref{successprob}). After as few as three measurements the concurrence is
essentially unity, indicating that the system is already in the desired
singlet state. The scheme does, however, not work with unit probability: The
probability $P_{M}$ of having projected system $A$ on the desired $\sigma
_{x}$ eigenstate converges to $\sim 0.2$ for the values above, a value
reached for $M=3$.

As another special case we consider a simplified form of the \textit{d}-wave
grain boundary qubit \cite{Blais:00}:$\ H=\Delta
(X_{1}+X_{2})+JZ(Z_{1}+Z_{2})$ ($\Delta $ is a tunneling parameter and $J$
is the Josephson coupling). Thus $\overrightarrow{A}=2(\Delta ,0,JZ)$, and
we may use the general result (\ref{eq8}), replacing $A_{y}$ by $A_{z}$. \
If we now choose $|\phi \rangle $ to be a $\sigma _{z}$-eigenstate then $%
V_{B}$ turns out to be unitary and hence all its eigenvalues have modulus
one, and no entanglement is generated. However, if we choose $|\phi \rangle $
to be a $\sigma _{x}$-eigenstate, we find 
\begin{eqnarray*}
V_{B} &=&\cos ^{2}\phi -\sin ^{2}\phi (\cos ^{2}\theta X_{1}X_{2}+\sin
^{2}\theta Z_{1}Z_{2}) \\
&&-\frac{i}{2}\sin 2\phi \cos \theta (X_{1}+X_{2})
\end{eqnarray*}%
where $\phi =\tau \sqrt{\Delta ^{2}+J^{2}}$ and $\tan \theta =J/\Delta $. It
is again easy to check that $V_{B}|s\rangle =|s\rangle $. One of the other three
eigenvalues is $1-2\sin ^{2}\phi \sin ^{2}\theta .$ The other two are
complex conjugates and for short times $\tau $ such that $\tan ^{2}\phi
<4\cos ^{2}\theta /\sin ^{4}\theta $ have the same amplitude $(1-2\sin
^{2}\phi \sin ^{2}\theta )^{1/2}$. For instance, when $\theta =\pi /4,$ the
amplitudes of the three eigenvalues are $\cos ^{2}\phi ,\cos \phi
,\cos \phi$, which vanish rapidly when raised to the power of the number of measurements.

Finally, note that the results presented so far are not restricted to $A$
being a qubit: the unitary operator (\ref{eq8}) is valid even when $A$ is a
multi-level system, as long as the condition $%
[A_{x},A_{y}^{2}]=[A_{y},A_{x}^{2}]=0$ is satisfied.

\subsection{Heisenberg model}

We now consider the Heisenberg interaction $H=J\overrightarrow{\sigma }\cdot
(\overrightarrow{\sigma }_{1}+\overrightarrow{\sigma }_{2})$. After showing
that $(H+J\overrightarrow{\sigma }_{1}\cdot \overrightarrow{\sigma }%
_{2})^{2}=9J^{2}$ and $[H,J\overrightarrow{\sigma }_{1}\cdot \overrightarrow{%
\sigma }_{2}]=0$, it is simple to compute $e^{-i\tau (H+J\overrightarrow{%
\sigma }_{1}\cdot \overrightarrow{\sigma }_{2})}$, from which we directly
obtain

\begin{eqnarray*}
U_{AB} &=&e^{-i\tau H}=e^{i\tau J\overrightarrow{\sigma }_{1}\cdot 
\overrightarrow{\sigma }_{2}}\{\cos (3\tau J) \\
&&-i\sin (3\tau J)[\overrightarrow{\sigma }\cdot (\overrightarrow{\sigma }%
_{1}+\overrightarrow{\sigma }_{2})+\overrightarrow{\sigma }_{1}\cdot 
\overrightarrow{\sigma }_{2}]/3\}.
\end{eqnarray*}%
$V_{B}$ then is identical to $U_{AB}$ provided one replaces with $%
\left\langle \phi \right\vert $ $\overrightarrow{\sigma }\left\vert \phi
\right\rangle $ the operator $\overrightarrow{\sigma }$ in $U_{AB}$. The
four eigenvalues are $\{1,e^{-2i\tau J},e^{i\tau J}\left( \cos 3\tau J\pm i%
\frac{1}{3}\sin 3\tau J\right) \}$, the last two having the same amplitude $%
1-\frac{8}{9}\sin ^{2}3\tau J<1$. The corresponding eigenvectors are,
respectively, $\vec{\beta}$ [Eq.~(\ref{eq:beta})]. Since the first two eigenvalues both have
magnitude $1$, we have the case discussed above:\ a pure entangled state
cannot be projected out by our method, but we can prepare a mixed entangled
state that is useful for all QIP\ protocols. Note further that these
Heisenberg model results hold for the entire class of Hamiltonians $H=%
\overrightarrow{A}_{\sigma }\cdot (\overrightarrow{\sigma }_{1}+%
\overrightarrow{\sigma }_{2})$, when $\overrightarrow{A}_{\sigma }$ is
generated from an arbitrary $2$-dimensional unitary transformation $U_{A}$, $%
\overrightarrow{A}_{\sigma }=U_{A}J$ $\overrightarrow{\sigma }U_{A}^{\dagger
}$ (this is different from the invariance under a rotation of the $B$ system
considered above).

\subsection{Bosonic media}

We now consider a photon or phonon ($A$) interacting symmetrically with two
identical qubits ($B$).\ The Jaynes-Cummings Hamiltonian is%
\begin{equation*}
H=\epsilon b^{\dagger }b+g(\sigma _{1}^{z}+\sigma _{2}^{z})+\frac{J}{2}%
[b(\sigma _{1}^{+}+\sigma _{2}^{+})+b^{\dagger }(\sigma _{1}^{-}+\sigma
_{2}^{-})]
\end{equation*}%
where $b$ is a bosonic annihilation operator, and whence $\overrightarrow{A}%
=(J(b+b^{\dagger }),iJ(b-b^{\dagger }),g)$. Such a model can easily be
realized using microwave cavity QED \cite{Rauschenbeutel:99} (two atoms in
one or two cavities) or trapped ions \cite{Wineland:98}. We can exactly
diagonalize $H$ by writing it as a direct sum over three-dimensional matrices in the basis $%
\left\vert n-1,t_{+}\right\rangle ,$ $\left\vert n,t_{0}\right\rangle $ and $%
\left\vert n+1,t_{-}\right\rangle $, where $\left\vert n\right\rangle $ are
number states and $\left\vert t_{\alpha }\right\rangle $ are triplet states
of the two qubits. Note that $\mathcal{N}\equiv \lbrack b^{\dagger
}b+(\sigma _{1}^{z}+\sigma _{2}^{z})]$ is a conserved quantity and hence $H$
is block diagonal in its eigenvalues. For simplicity we consider the case
with $g=\epsilon $ and measure the single photon state \ $\left\vert \phi
\right\rangle =b^{\dagger }\left\vert 0\right\rangle =|1\rangle $ (this
projects onto a single block of the infinite matrix $U_{AB}$). In this case
we readily find the already diagonal 
\begin{eqnarray*}
V_{B}(\tau ) &=& \mathrm{diag}\{1,e^{-2i\epsilon \tau }\frac{3+2\cos (\sqrt{10}%
  \tau J)}{5},\\
&e^{-i\epsilon \tau }&\cos (\sqrt{6}\tau J),\cos (\sqrt{2}\tau J)\}
\end{eqnarray*}
in the ordered basis $\{|1\rangle |s\rangle ,\left\vert 1,t_{+}\right\rangle
,\left\vert 1,t_{0}\right\rangle ,\left\vert 1,t_{-}\right\rangle \}$. Thus,
as long as we make sure that $\tau <2\pi /\sqrt{10}J$ the method will
project out a pure singlet state. It follows from our discussion of
invariance above that another Bell state $\frac{1}{\sqrt{2}}(|01\rangle
+|10\rangle )$ can be projected out if the two qubits couple to the photon
or phonon with opposite signs.

\subsection{Multilevel Systems}

Let us now consider a rather general, though abstract multilevel case. Assume that
subsystem $B$ consists of two particles that have $M\geq 2$ levels each, and
that the interaction Hamiltonian with subsystem $A$ (also an $M$-level
system) is of the form

\begin{equation*}
H=\sum_{i,j=1}^{M}A_{ij}(O_{ij}^{1}+O_{ji}^{2}),
\end{equation*}%
where $O_{ij}=E_{ij}-E_{ji}$ and $E_{ij}=\left\vert i\right\rangle
\left\langle j\right\vert $ is a matrix whose elements are zero everywhere
except for a $1$ at position $(i,j)$. Namely, there is an $SO(M)$ symmetry
for odd $M$ or an $Sp(M)$ symmetry for even $M$. In this case $H$ has a 
\emph{single} non-degenerate eigenvector with \emph{zero} eigenvalue. This
state is entangled. For instance, when $N=3$, we have two qutrits, which can
be represented in the spherical basis $\{\left\vert -1\right\rangle
,\left\vert 0\right\rangle ,\left\vert +1\right\rangle \}$. The state with
zero eigenvalue is 
\begin{equation*}
|\Psi \rangle =(\left\vert 1\right\rangle _{1}\left\vert -1\right\rangle
_{2}-\left\vert 0\right\rangle _{1}\left\vert 0\right\rangle _{2}+\left\vert
-1\right\rangle _{1}\left\vert 1\right\rangle _{2})/\sqrt{3},
\end{equation*}%
and is maximally entangled. In general, the one-dimensional subspace
containing the state with zero eigenvalue is the irreducible representation $%
(0,0,...,0)$ of $SO(M)$ or $Sp(M)$. This state is the generalization of the
singlet state that arose in the general theory and examples treated above
when the subsystems were qubits.

\section{Perturbative treatment}

It is apparent from the above examples that the success of our method
depends on keeping the period $\tau $ between two measurements short. For
instance, in the bosonic example, we require $\tau <2\pi /\sqrt{10}J$ in
order to prevent the appearance of another eigenvalue with amplitude one.
Let us therefore consider a short-time expansion of $V_{B}(\tau )$. To first
order
\begin{eqnarray*}
V_{B}(\tau )^{M} &=& (I-i\tau \left\langle \phi \right\vert
H_{AB}\left\vert \phi \right\rangle )^{M} \\
&\overset{M\rightarrow \infty }{%
\longrightarrow }& \exp (-it\left\langle \phi \right\vert H_{AB}\left\vert
\phi \right\rangle ),
\end{eqnarray*}
where $M\tau =t$ (constant), i.e., the evolution is
unitary. This is the quantum Zeno effect, which completely decouples the
interaction among all parties of subsystem $B$, so that no entanglement can
be generated in this limit. Thus the dominant contribution to entanglement
generation originates from the second-order term, which contains the \emph{%
self-correlation} of subsystem $B$. Consider for simplicity the case $%
\left\langle \phi \right\vert H_{AB}\left\vert \phi \right\rangle =0$ (as in
our last example). Letting $M\tau ^{2}/2=t^{2}$ (constant), we have 
\begin{equation*}
V_{B}(\tau )^{M}\overset{M\rightarrow \infty }{\longrightarrow }%
e^{-t^{2}\left\langle \phi \right\vert H_{AB}^{2}\left\vert \phi
\right\rangle }
\end{equation*}%
All eigenstates of $\left\langle \phi \right\vert H_{AB}^{2}\left\vert \phi
\right\rangle $ with nonzero eigenvalues are rapidly suppresed as $t$ (in
practice $M$) increases, while those with eigenvalue zero survive. Since it
is much simpler to analytically calculate $\left\langle \phi \right\vert
H_{AB}^{2}\left\vert \phi \right\rangle $ than $V_{B}(\tau )$, this
perturbative method provides a relatively simple tool for estimating the
possibility of entanglement generation via our method, for complicated
systems. We consider spin-orbital coupling as another illustrative example: $%
H=hL_{x}(X_{1}+X_{2})+hL_{y}(Y_{1}+Y_{2})$, denoting two spin qubits that
couple with the same orbital angular momentum. If we measure the eigenstate $%
\left\vert \phi \right\rangle =\left\vert l,0\right\rangle $ of the orbital
component, we find $\left\langle \phi \right\vert H\left\vert \phi
\right\rangle =0$ and $\left\langle \phi \right\vert H^{2}\left\vert \phi
\right\rangle =hl(l+1)(2+X_{1}X_{2}+Y_{1}Y_{2})$. The eigenvalues are $%
hl(l+1)\{0,2,4,2\}$ with respective eigenvectors $\vec{\beta}$, meaning that
the singlet state $\left\vert s\right\rangle $ is projected out in the
second-order analysis. Note that this example is qualitatively different
from the previous one since subsystem $A$ here does not refer to a
physically separate particle.

\section{Preparation of long-distance entanglement}

Finally we come to our main result: the generation of long-distance
bi-partite entanglement. Since in our discussion above there was no
restriction on the size of the $A$ subsystem, in a sense long-distance
entanglement generation already follows from the results
above. However, it is important to specify how the measurements on $A$
can be carried out. Assume that the $A$ subsystem is composed of $N-2$
qubits $2,3,...,N-1$ and we wish to entangle the two $B$-subsystem qubits $%
1,N$. We consider again as an illustrative example a simplified form of the 
\textit{d}-wave grain boundary qubit \cite{Blais:00}: 
\begin{widetext}
\begin{eqnarray*}
U_{AB} =\exp [-iJ\tau (X_{1}+X_{N}) 
-i\Delta \tau (Z_{1}Z_{2}+Z_{2}Z_{3}+h_{3}+Z_{N-2}Z_{N-1}+Z_{N-1}Z_{N})]
\end{eqnarray*}%
where $h_{3}=\sum_{i=3}^{N-3}Z_{i}Z_{i+1}$. We choose $|\phi \rangle $ to be
the state $\left\vert +\right\rangle _{2}\left\vert R\right\rangle
\left\vert +\right\rangle _{N-1}$, where $\left\vert R\right\rangle
=\left\vert +\right\rangle _{3}\left\vert +\right\rangle _{4}...\left\vert
+\right\rangle _{N-2}$, where $|+\rangle =(|0\rangle +|1\rangle )/\sqrt{2}$.
It follows after some calculations that 
\begin{eqnarray*}
V_{B} &=&\frac{1}{2}\{a[\cos ^{2}\phi -\sin ^{2}\phi (\cos ^{2}\theta
X_{1}X_{N}+\sin ^{2}\theta Z_{1}Z_{N})-i\frac{1}{2}\sin 2\phi \cos \theta
(X_{1}+X_{N}) \\
&&+b[\cos ^{2}\phi -\sin ^{2}\phi (\cos ^{2}\theta X_{1}X_{N}-\sin
^{2}\theta Z_{1}Z_{N})-i\frac{1}{2}\sin 2\phi \cos \theta (X_{1}+X_{N})\}
\end{eqnarray*}%
where 
\begin{eqnarray*}
a &=&\left\langle R\right\vert e^{-i\Delta \tau
(Z_{3}+Z_{N-2}+h_{3})}\left\vert R\right\rangle =\left\langle R\right\vert
e^{-i\Delta \tau (-Z_{3}-Z_{N-2}+h_{3})}\left\vert R\right\rangle  \\
b &=&\left\langle R\right\vert e^{-i\Delta \tau
(Z_{3}-Z_{N-2}+h_{3})}\left\vert R\right\rangle =\left\langle R\right\vert
e^{-i\Delta \tau (-Z_{3}+Z_{N-2}+h_{3})}\left\vert R\right\rangle 
\end{eqnarray*}
\end{widetext}
(and are real numbers for odd $N$), $\phi =\tau \sqrt{J^{2}+\Delta ^{2}}$, $%
\tan \theta =\Delta /J$. Two of the eigenvalues and eigenstates are:

\begin{eqnarray*}
E_{1} &=&\frac{1}{2}\{a+b(1-2\sin ^{2}\phi \sin ^{2}\theta )\};\frac{1}{%
\sqrt{2}}(\left\vert 1_{1}0_{N}\right\rangle -\left\vert
0_{1}1_{N}\right\rangle ), \\
E_{2} &=&\frac{1}{2}\{b+a(1-2\sin ^{2}\phi \sin ^{2}\theta )\};\frac{1}{%
\sqrt{2}}(\left\vert 0_{1}0_{N}\right\rangle -\left\vert
1_{1}1_{N}\right\rangle ).
\end{eqnarray*}%
Thus, as desired, entanglement is generated between the $B$ subsystem
particles $1,N$. Note that since the measurement of all the particles of the 
$A$ subsystem is carried out in parallel \emph{this method for long-range
entanglement generation is independent of the distance} $N$.

The amplitudes of the other two eigenvalues are $\sqrt{E_{1}E_{2}}$ in the
case of odd $N\geq 5$. Since $\max (|E_{2}|,|E_{1}|)>\sqrt{E_{1}E_{2}}$ for
sufficiently short time, and $E_{1}-E_{2}=2\left( \sin \phi \right)
^{2}\left( \sin \theta \right) ^{2}\left( a-b\right) $ as long as $a\neq b$,
the procedure prepares either $\frac{1}{\sqrt{2}}(\left\vert
1_{1}0_{N}\right\rangle -\left\vert 0_{1}1_{N}\right\rangle )$ or $\frac{1}{%
\sqrt{2}}(\left\vert 0_{1}0_{N}\right\rangle -\left\vert
1_{1}1_{N}\right\rangle )$. We find, e.g, for $N=5$: $a=\cos 2\Delta \tau $, 
$b=1$ and $\frac{1}{\sqrt{2}}(\left\vert 00\right\rangle -\left\vert
11\right\rangle )$ is projected out, while for $N=7$: $a=\frac{1}{4}(3+\cos
4\Delta \tau )$, $b=\cos 2\Delta \tau $ and $\frac{1}{\sqrt{2}}(\left\vert
01\right\rangle -\left\vert 10\right\rangle )$ is projected out. The case of
even $N$ is more complicated since the amplitudes of other two eigenvalues
depend on the values of $a$ and $b$. E.g., in the $N=4$, $V_{B}$ has the
same form as above but $a=e^{-i\Delta \tau }$ and $b=e^{i\Delta \tau }$; all
roots have the same amplitude and no pure state will be projected out.$%
\allowbreak $

\section{Conclusions}

We have introduced a measurement-based method for entangling two systems
that only interact indirectly, via a third system. Applications range from
solid state to atomic quantum information processing. One of the advantages
of this measurement-based method of entanglement generation is that it does
not depend on precisely timed interactions, as are interactions-based
schemes. The method can be used to prepare arbitrarily long-distance
entanglement via a chain of intermediate particle in one parallel
measurement step. This may have useful applications in reducing the latency
overhead in quantum communication in quantum computing architectures. An
interesting open question which we leave for future research is whether the
same method can be used to apply quantum logic gates between non-interacting
particles.

Acknowledgents.--- D.A.L. acknowledges support from the Sloan
Foundation, D-Wave Systems, Inc., and the
DARPA-QuIST program (managed by AFOSR under agreement
No. F49620-01-1-0468).


\end{document}